\documentclass[notitlepage,amsmath, amssymb]{revtex4-1}

\pdfoutput=1

\usepackage{tabularx}
\usepackage{graphicx}

\newcommand{\chem}[1]{\mbox{\tiny #1}}
\newcommand{\oSE}{\mathrm{1SE}}

\begin{document}

\title{Sparse Modeling analysis of Extended X-ray Absorption Fine Structure data using two-body expansion}

\author{Fabio Iesari$^{1,2}$, Hiroyuki Setoyama$^2$, Yasuhiko Igarashi$^{3,4,5}$, Masato Okada$^{3,5}$, Hiroyuki Kumazoe$^6$, Kazunori Iwamitsu$^7$, Ichiro Akai$^{6,2}$, Yoshiki Seno$^2$ and Toshihiro Okajima$^{1,2}$}

\affiliation{$^1$Aichi Synchrotron Radiation Center, Seto, Aichi, 489-0965, Japan
\\
$^2$Kyushu Synchrotron Light Research Center, Tosu, Saga, 841-0005, Japan
\\
$^3$Graduate School of Frontier Science, The University of Tokyo, Kashiwa, Chiba, 277-8561, Japan
\\
$^4$Japan Science and Technology Agency, PRESTO,  Kawaguchi, Saitama, 332-0012, Japan
\\
$^5$Research and Services Division of Materials Data and Integrated System, National Institute for Material Science, Tsukuba, Ibaraki, 305-0047, Japan
\\
$^6$Institute of Industrial Nanomaterials, Kumamoto University, Kumamoto, 860-8555, Japan
\\
$^7$Technical Division, Kumamoto University, Kumamoto, 860-8555, Japan
}

\date{\today}

\begin{abstract}
Analysis of extended X-ray absorption fine structure (EXAFS) data by the use of sparse modeling is presented. We consider the two-body term in the $n$-body expansion of the EXAFS signal to implement the method, together with calculations of amplitudes and phase shifts to distinguish between different back-scattering elements. Within this approach no \textit{a priori} assumption about the structure is used, other than the elements present inside the material. We apply the method to the experimental EXAFS signal of metals and oxides, for which we were able to extract the radial distribution function peak positions, and the Debye-Waller factor for first neighbors.
\end{abstract}

\maketitle

\section{Introduction}
\label{sec:int}

X-ray absorption spectroscopy is now an established experimental technique widely used in materials science and many other scientific fields \cite{Cal2013,Bun2010}. In the energy region beyond a core-electron edge, oscillations in the absorbance are present, which are usually referred to as XAFS (X-ray absorption fine structure). The excited photo-electron is back-scattered by the potential of neighboring atoms, resulting in an interference that modulates the absorption cross-section, depending both on the incoming energy of the x-rays and on the inter-atomic distances in the material investigated. As a consequence, analysis of these oscillations permits to extract information on the local structure around the photo-absorbing atom.

The oscillations that continues after 50 eV above the absorption edge are commonly known as EXAFS (extended X-ray absorption fine structure). The EXAFS structural signal is defined as the fractional oscillations in the absorption respect to the atomic background and normalized to the atomic cross-section of the edge under consideration, $\chi(k) = (\mu(E)-\mu_0 (E))/\mu_{\chem{edge}}(E)$, where $k = \sqrt{2m(E-E_0)} / \hbar$ is the photo-electron wavevector. The EXAFS signal can be decomposed in terms of the $n$-body distribution functions \cite{FilDic1995a}, which are statistical quantities that describe the structure of a material \cite{AllTil1987}:

\begin{equation}
\chi(k) = \int_0^\infty dr 4 \pi r^2 \rho g_2(r) \gamma^{(2)}(r,k) + \mathcal{O}(g_3)
\label{eq:xas}
\end{equation}

The direct inversion of this integral equation is known to be an ill-posed problem, so conventional analysis methods make use of assumptions or additional information to interpret the experimental data. The most common approach, so called peak-fitting scheme, consists in modeling the structure with a given number of peaks dependent on a set of parameters and then refine those parameters using a least-square method to fit either the Fourier Transform (FT) of the signal \cite{RavNew2005} or the raw experimental data \cite{FilDic1995b}. Prior knowledge of the structure is required for choosing the peaks and calculate their signals. A more complex method, Reverse Monte Carlo (RMC) \cite{IesHat2018, TimKuz2012}, requires knowledge of the density and chemical composition of the system and, since it is a simulation method, it is also costly in terms of time and computing power. More recently, a method based on Neural Networks (NN) \cite{TimAns2018, TimAhm2019} has been successfully applied to the analysis of EXAFS, although this requires a specialized trained NN case by case.

To overcome these limitations, an alternative approach that does not require prior knowledge about the structure is to use sparse modeling \cite{AkaIwa2018}. This method falls within the so-called regularization technique \cite{BabVas1981}: in our case we utilize the $\ell_1$-norm as regularization condition, which tends to find sparse solutions \cite{Tib1996}, hence the name sparse modeling. This is because, in general, the $L_1$-regularizer prioritize relevant features and keep the others to zero to avoid overfitting of the data. In previous implementation of the method \cite{AkaIwa2018, SetAka2020}, simple sine and cosine functions were used as basis to describe the oscillations of the EXAFS signals together with few parameters and could obtain Debye-Waller factors for various metals. In this paper, we show progress of this method using a more refined model: starting from the two-body expansion of the EXAFS signal and including calculations for amplitudes and phase shifts, we are able to extract the peak positions of the RDF and to treat multi-atomic cases.

The idea is to have an analysis method which is almost completely automated and fast for quick check of data files during experiments and that can be used also by people who may not be familiar with standard analysis procedures.

\section{Method}
\label{sec:met}

For sparse modeling we use a least absolute shrinkage and selection operator (LASSO) \cite{SanSym1986,Tib1996} which uses $\ell_1$ regularization. Within this method, a linear mapping of the type $\vec{y}  = {\bf M} \vec{x}$ is solved for $\vec{x}$, where $\vec{y}$ is the target data, the EXAFS signal in our case, $\vec{x}$ is the solution vector and $\bf M$ is a matrix whose elements are defined by the problem at hand. This is obtained by solving the following regularized least square problem: 
\begin{equation}
\hat{\vec{x}}(\lambda) = \arg \min_{\vec{x}} \left( \| \vec{y} - M \vec{x} \|_2^2 + \lambda \| \vec{x} \|_1 \right)
\label{eq:lasso}
\end{equation}
the first quantity on the r.h.s. of the equation represents the solution to the linear problem, while the second term is the $\ell_1$-norm of the solution vector: the hyperparameter $\lambda$ controls the sparsity of the solution by weighting the $\ell_1$-norm term. In order to define the elements of the matrix ${\bf M}$ for our analysis, we need an appropriate description of the EXAFS signal.

The EXAFS signal $\chi(k)$ can be expanded in terms of the $n$-body distribution functions \cite{FilDic1995a}, where the main contribution is given by the two-body term, as written in Eq.~\eqref{eq:xas}. The $\gamma^{(2)}$ are the irreducible two-body signals, corresponding to the presence of an atom at distance $r$ from the photo-absorber. Each signal $\gamma^{(2)}(r,k)$ is usually written as an oscillating function of the type $A(k,r)\sin[2kr+\phi(k,r)]$. In this definition, the amplitude $A(k,r)$ contains not only the back-scattering amplitude, but also the mean free path effect and the term $1/kr^2$ \cite{Fil1994}. The phase shift $\phi(k,r)$ is the cause of the mismatch between the position of the peaks in the FT of $\chi(k)$ and the actual atomic distances in the structure. The $g_2(r)$ is the radial distribution function (RDF) and is defined as the probability of finding two atoms at a distance $r$ \cite{AllTil1987}: it therefore contains information on the structure of the system. Sometimes in literature with the same symbol $g(r)$ is indicated the same quantity with the inclusion of the volume term $4\pi r^2 \rho$. For the sake of clarity, we will differentiate the two by using $n(r) = 4 \pi r^2 \rho g(r)$ \cite{Was1980}, dropping also the subscript for short. This distinction becomes important when treating multi-component systems.

We approximate the RDF $n(r)$ as a vector by binning it with a given interval $\Delta r$: within this approximation we transform the integral from Eq.~\eqref{eq:xas} into a Riemann sum of rectangles:
\begin{equation}
\chi(k) \simeq \sum_{i=1}^N n(r_i) A(k,r_i) \sin[2kr_i + \phi(k,r_i)] \Delta r
\label{eq:xas2}
\end{equation}
where the sum goes from $r_1 = 1.5$ {\AA} to a cut-off $r_N$ because of the short-range nature of EXAFS. From Eq.~\eqref{eq:xas2} we define then our LASSO problem as follows: $n(r_i)$ is the solution $\vec{x}$, $\vec{y}$ is the $k$-weighted experimental signal $k^w \chi(k)$ and the matrix is composed by the elements:
\begin{equation}
M_{ji} = A(k_j,r_i)\sin[2 k_j r_i+\phi(k_j,r_i)] \Delta r
\label{eq:matrix}
\end{equation}

The appropriate $k$-weight $w$ for the fitting is chosen case by case (typically $w = 1$-$3$), depending on the signal intensity at high $k$, if oscillations are still visible, and the data noise. In this work we used $w = 2$ for the metal samples and $w = 1$ for copper oxide. The amplitudes and phase shifts for all the distances $r_i$ considered are calculated with GNXAS \cite{redbook, gnxas}, which uses a complex Hedin-Lundqvist potential model \cite{HedLun1971}, that includes inelastic loss effects, and muffin-tin approximation, where the radii are chosen according to Norman's criterion \cite{Nor1976}. In the EXAFS region, above $k \simeq 4$ {\AA$^{-1}$}, these approximations can be used without much loss of accuracy: as a consequence, atomic scattering properties are independent of atomic positions \cite{Fil2001} and the signals in our scheme can be calculated even if the structure is not known, apart from which elements are present. 

In Eq.~\eqref{eq:xas2} we also did not include the reduction factor $S_0^2$, which accounts for the many-body corrections to the one-electron cross section \cite{FilDic1995b} and usually lays in the interval 0.7-1.0. Because $S_0^2$ is just a multiplication factor for the signal, its inclusion in the regularization scheme would make little sense, since it would correlate with the amplitude of the solution vector and the hyper-parameter $\lambda$. This is equivalent to set $S_0^2 = 1$.

Although the solution vector is set as $n(r)$, because of the lack of three-body components and of the $S_0^2$ reduction factor, a complete reconstruction of the RDF is not possible, but still we can obtain peak positions to have insights about the structure. Additionally, this method is not to be used for disordered systems, as it has been demonstrated that without the use of proper additional constraints on the RDF, we can achieve a good fit with erroneous results \cite{Fil2001}.

Up to now we have treated the mono-atomic case for simplicity, but the method is extendable to multi-atomic systems by substituting the total RDF with the partial RDF (pRDF). In this case Eq.~\eqref{eq:xas2} becomes:

\begin{equation}
\chi_{\alpha}(k) \simeq \sum_{\beta} \sum_{i=1}^N n_{\alpha \beta}(r_i) A_{\alpha \beta}(k,r_i) \sin[2kr_i + \phi_{\alpha \beta}(k,r_i)] \Delta r
\label{eq:multi}
\end{equation}
the signal measured at the edge of element $\alpha$ is proportional to the partial RDFs of $\alpha$ respect to all elements present in the system ($\alpha$ itself included) \footnote{Note that $n_{\alpha \beta}(r)$ is not symmetric for inversion of atoms, meaning $n_{\alpha \beta}(r) \neq n_{\beta \alpha}(r)$, while the $g(r)$ is, $g_{\alpha \beta}(r) = g_{\beta \alpha}(r)$; this is why it is important to distinguish the two.}. The vector $\vec{x}$ contains the different partial RDFs in succession \[\vec{x} = \left[ n_{\alpha \alpha}(r_1), \ldots, n_{\alpha \alpha}(r_N), n_{\alpha \beta}(r_1), \ldots,n_{\alpha \beta}(r_N), n_{\alpha \gamma}(r_1), \ldots \right] \] and the LASSO matrix $\bf M$ is expanded and calculated accordingly.

To choose the appropriate value of the hyper-parameter $\lambda$ five-fold cross validation analysis and the one-standard-error (1SE) rule are employed \cite{HasTib2009,Mur2012,AkaIwa2018}. We divide the data into five different subset randomly and then iteratively use each subset to test the result obtained from applying LASSO to the remaining data points. For all values of $\lambda$ in a given interval, the root mean square deviation (RMSD) between the model and the selected data subset is calculated. For each $\lambda$, the cross validation error (CVE) is defined as the average between the RMSD of the five subset and the standard error (SE) as its standard deviation. The 1SE rule then chooses the largest $\lambda$ among those whose error is within one standard error of the minimum CVE. The optimized $\lambda$ is denoted by $\lambda_{\oSE}$. Since cross validation (CV) datasets are chosen randomly each time, we can repeat the procedure different times and consider the average as the result.

\section{Results and Discussion}
\label{sec:res}

The method is implemented using Python3 \cite{python} and the \texttt{scikit-learn} package \cite{scikit} for the LASSO subroutines. All spectra were collected at the beamline BL11 \cite{OkaSum2013} of the Saga Light Source using transmission geometry. The samples are references available at the beamline, consisting in thin films for metals and a pellet obtained from a powder mixture for copper oxide. The EXAFS signals $\chi(k)$ are extracted from the raw experimental data using the program \texttt{jesf} from the GNXAS package \cite{redbook}, which consists in removing the atomic background, fitting the post-edge background with a smooth polynomial function and normalize the spectra. The edge energy is selected as the energy point on the rising edge where the derivative of the absorption is maximum.

\subsection{Cu}

\begin{figure}[t]

\centering
\includegraphics[width=0.99\textwidth]{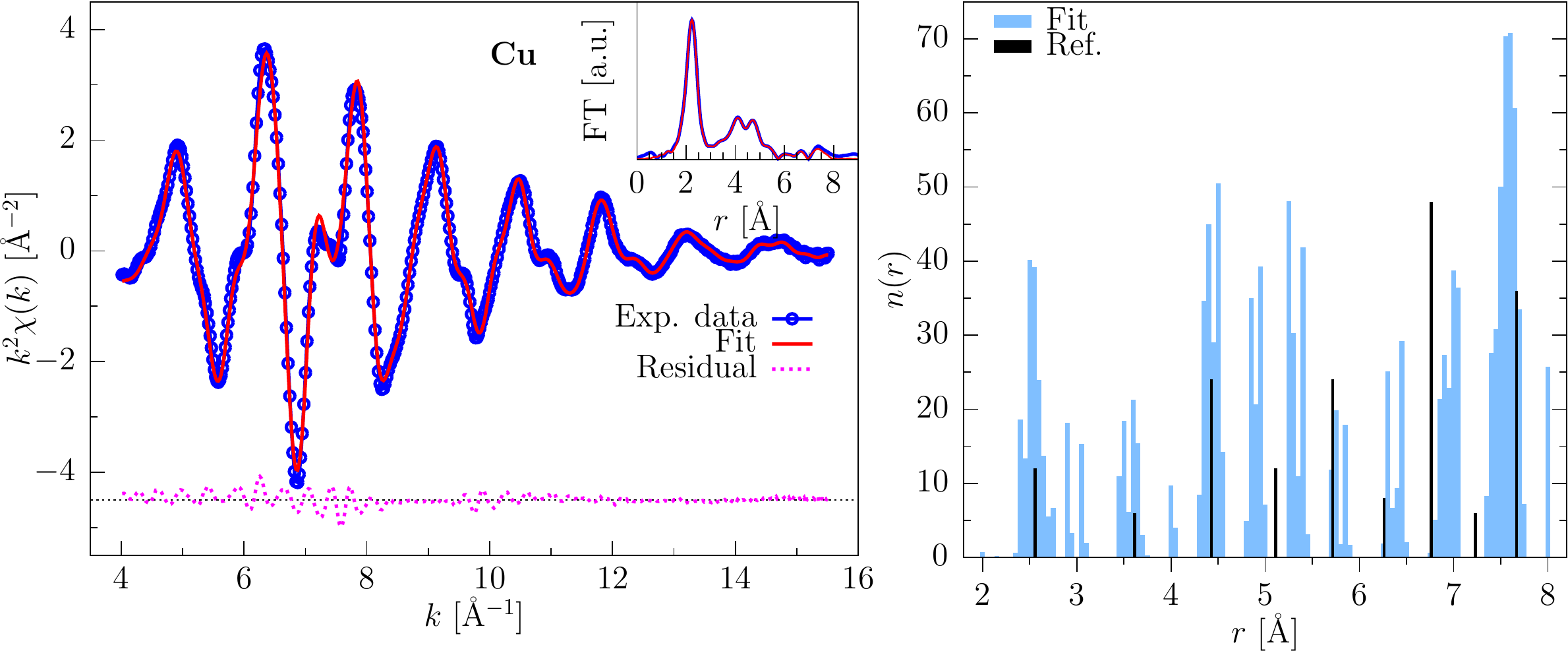}

\caption{\label{fig:cu1}Results of the LASSO procedure for copper K-edge spectrum. Left: the experimental EXAFS signal is compared with the fit, FTs are shown in the inset. Right: obtained RDF is compared with the crystal peak positions.}

\end{figure}

 We will first consider the case of metal copper to discuss more details about the method and draw some conclusions from the result. In Fig.~\ref{fig:cu1} we show the experimental Cu K-edge EXAFS signal $k^2 \chi (k)$ together with the results of the LASSO procedure, averaged from four different CV datasets. The RDF is compared with the positions of the crystal shells scaled to their coordination number, taken from X-ray diffraction (XRD) \cite{Ott1961}, up to 8 {\AA}, which is the distance cut-off we used for the fitting. From the comparison, we can see that the obtained RDF has peaks in the correct positions in the range considered, with the exception of the peak at around 5.1 {\AA}. The reason can be attributed to the presence of a strong three-body signal with half-path length around that distance, coming from the collinear configuration of three first-neighbor copper atoms. Since the model is only considering two-body contributions, additional oscillations due to three-body distributions are fitted and results in spurious peaks, like the ones that appears around 3 and 4 {\AA}. These contributions are usually small, with the exception of collinear configurations due to the focusing effect \cite{RehAlb2000}. We have also to consider that because the intensity of EXAFS decreases very rapidly with the increasing of distances, due to mean free path effects, the degree of accuracy at higher distances decreases. This can be seen from the FT of Fig.~\ref{fig:cu1}, where signals above 6 {\AA} are relatively weak. Nonetheless we decided to include contributions up to 8 {\AA} because it is clear that oscillations are still present and it makes the numerical procedure more stable.

From the first peak we calculated the average distance $\bar{r}$ and the Debye-Waller factor $\sigma^2$ using the following definitions:
\begin{equation}
\bar{r}  =  \frac{\sum_i r_i n(r_i) \Delta r}{\sum_i n(r_i) \Delta r} 
\end{equation}
\begin{equation}
\sigma^2  =  \frac{ \sum_i (r_i - \bar{r})^2 n(r_i) \Delta r}{\sum_i n(r_i) \Delta r }
\end{equation}
where the index $i$ runs over the range of the peak. We calculated these values for each of the four different CV datasets and considered the average. From the result, it is also possible to calculate the coordination number, but because we did not include the reduction factor $S_0^2$ and because of $\ell_1$-norm penalty term in the LASSO Eq. \eqref{eq:lasso}, the solution coefficients are underestimated and so is the coordination number.

 In Table~\ref{tab:cu_param} we report the obtained parameters, compared with the same parameters obtained through the conventional peak-fitting scheme within GNXAS for the same data and literature references. Similar to previous works \cite{AkaIwa2018, SetAka2020}, the error on the Debye-Waller is the standard deviation of the mean over the CV datasets. The error on the average distance, instead, comes from a non-structural parameter, the energy shift $\Delta E$, which is the difference between the experimental energy scale and the energy scale of the phase shift and amplitude calculations. This difference is due to both practical and theoretical reason \cite{redbook} and usually lies in the interval 0-10 eV. This parameter is known to correlate with the inter-atomic distances $r$ \cite{FilDic1995b}, hence fixing it means that the average distances obtained is biased and the uncertainty is bigger than expected. In the calculations of amplitudes and phase-shifts the value was fixed at 4 eV, so for an uncertainty on $\Delta E$ of 5 eV, we estimated an expected error on the distances of about $\Delta r \simeq $  0.05 {\AA}, which is also consistent with our choice for binning the RDF.

The parameters obtained agrees well with standard fitting analysis and reference data. Without making any \textit{a priori} assumption about the structure, we were able to fit the experimental data accurately (Fig.~\ref{fig:cu1})  and obtain peak positions of the RDF, consistent with diffraction data. 

\begin{table}
\renewcommand{\arraystretch}{1.3}
\begin{tabularx}{0.6\columnwidth}{c
 >{\centering}X
 >{\centering\arraybackslash}X
 }
\hline
\hline
 & $r$ [{\AA}] & $\sigma$ [{\AA}] \\
\hline
SpM  & 2.54 $\pm$ 0.05 & 0.087 $\pm$ 0.006 \\
GNXAS & 2.54 $\pm$ 0.02 & 0.090 $\pm$ 0.010  \\
EXAFS\cite{TohUda1985} & 2.51 & 0.08 \\
Calc.\cite{SevMeu1979} & & 0.088 \\
\hline
\hline
 \end{tabularx}
\caption{\label{tab:cu_param}Parameters of the first peak of the RDF for Cu obtained by sparse modeling (SpM) and by peak-fitting scheme using GNXAS for the same experimental data. Values taken from references are also shown for comparison.} 
\end{table}

 \begin{figure}[t]

\centering
\includegraphics[width=0.99\textwidth]{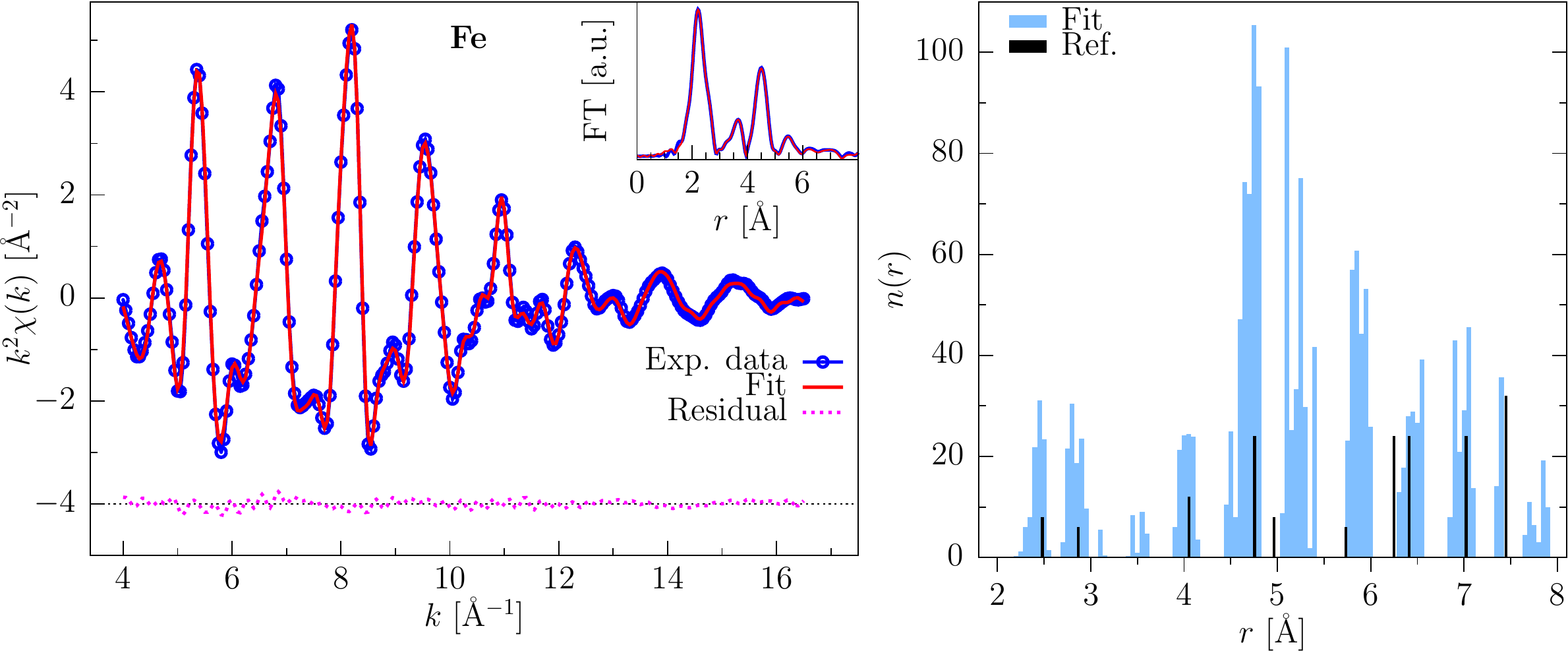}

\caption{\label{fig:fe}Results of the LASSO procedure for iron K-edge spectrum. Left: the experimental EXAFS signal is compared with the fit, FTs are shown in the inset. Right: obtained RDF is compared with the crystal peak positions.}

\end{figure}

\begin{table}
\renewcommand{\arraystretch}{1.3}
 \begin{tabularx}{0.6\columnwidth}{c
 >{\centering}X
 >{\centering\arraybackslash}X
 }
\hline
\hline
 & $r$ [{\AA}] & $\sigma$ [{\AA}] \\
\hline
&  (Peak I)   &\\
SpM  & 2.43 $\pm$ 0.05 & 0.063 $\pm$ 0.003 \\
GNXAS & 2.45 $\pm$ 0.02 & 0.059 $\pm$ 0.015  \\
Calc.\cite{SevMeu1979} & & 0.071 \\
\hline
& (Peak II)  & \\
SpM & 2.84 $\pm$ 0.05 & 0.087 $\pm$ 0.005 \\
GNXAS  & 2.83 $\pm$ 0.07 & 0.070 $\pm$ 0.028  \\
Calc.\cite{SevMeu1979}  & & 0.081 \\
\hline
\hline
 \end{tabularx}
 \caption{\label{tab:fe_param}Parameters of the first and second peak of the RDF for Fe obtained by sparse modeling (SpM) and by peak-fitting scheme using GNXAS for the same experimental data. Values of the Debye-Waller taken from calculations are also shown for comparison.}
  \end{table}

\subsection{Fe}

We applied the same method to the Fe K-edge of an iron foil. In Fig.~\ref{fig:fe} we report the experimental EXAFS signal compared with the fit and the obtained RDF compared with peak positions. Iron has a body centered cubic structure with a cell parameter $a_{\chem{Fe}} = 2.8665$ {\AA} \cite{Wyc1963}. The EXAFS signal in this case is rather complex, but it is fitted quite accurately. The conclusions we draw from the Cu case, can also be applied to the RDF obtained here. Peaks are correctly identified, in particular at short distance where the two distinct peaks for bbc structure appear. At longer distances agreement is not as accurate due to mean free path effect that makes the EXAFS signal from these distant shells weak. In Table~\ref{tab:fe_param} we report the average distance and Debye-Waller factors obtained for the first two peaks. Again, the accuracy on the average distance comes from the fix value of the energy shift $\Delta E$, while for the Debye-Waller factor is the standard deviation of the mean across four CV datasets. We can see that values obtained agrees with the values obtained from the conventional EXAFS analysis of the same data. 

\begin{figure}[t]

\centering
\includegraphics[width=0.99\textwidth]{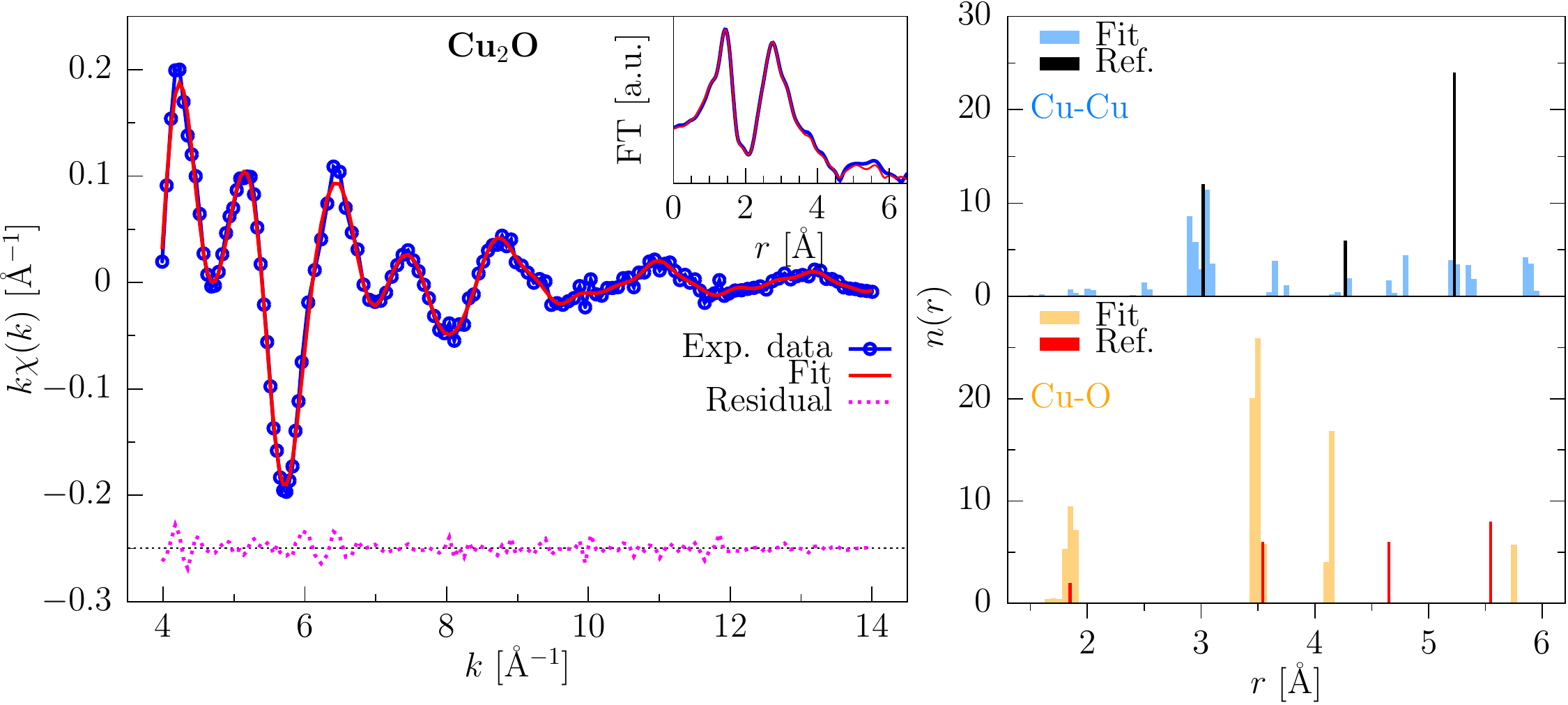}

\caption{\label{fig:oxides}Results of the LASSO procedure for cuprous oxide Cu K-edge spectrum. Left: the experimental EXAFS signal is compared with the fit, FTs are shown in the inset. Right: obtained partial RDFs, compared with atomic peak positions taken from crystallographic data.}

\end{figure}

\subsection{Cu$_2$O}
 
As explained in Sect.~\ref{sec:met}, the method is expanded to multi-atomic systems by taking into consideration the partial RDFs. To test a multi-atomic case, we measured absorption spectra of cuprous oxide, Cu$_2$O, at the K-edge of Cu. Since EXAFS is chemically selective, we are able to probe only RDF where one of the element involved corresponds to the edge we are measuring; therefore in this case, we are sensitive to the two partial RDF $n_{\chem{Cu-Cu}}(r)$ and $n_{\chem{Cu-O}}(r)$, while being completely blind to $n_{\chem{O-O}}(r)$. Results of the LASSO calculations are shown in Fig.~\ref{fig:oxides}. The pRDF are compared with crystal peaks position scaled to their coordination numbers \cite{KirEic1990} up to 6 {\AA}, which is the cut-off used for this case. The $k$-weight on the EXAFS signal used in this case is $w = 1$, because of the more noisy data. The experimental signal in this case presents few oscillations and it is therefore fitted by a small amount of features in the RDFs. Positions of the first three peaks are correctly identified, around 1.85 {\AA} and 3.5 {\AA} for Cu-O and at 3 {\AA} for Cu-Cu, and they contribute to almost the totality of the signal. In Table~\ref{tab:cu20_param}, we report the parameters of the first peak in the Cu-O pRDF, which gives the main oscillation to the signal. Similarly to the previous cases, the values for the average distance and Debye-Waller factor agrees with values obtained by standard analysis using GNXAS and reference data.

 \begin{table}
\renewcommand{\arraystretch}{1.3}
 \begin{tabularx}{0.7\columnwidth}{c
 >{\centering}X
 >{\centering\arraybackslash}X
 }
\hline
\hline
 & $r$ [{\AA}] & $\sigma$ [{\AA}] \\
\hline
 & (Cu-O)  &\\
SpM  & 1.85 $\pm$ 0.05 & 0.051 $\pm$ 0.004 \\
GNXAS  & 1.86 $\pm$ 0.02 & 0.058 $\pm$ 0.035  \\
EXAFS\cite{TohUda1985} & 1.85 & 0.04 \\
\hline
\hline
 \end{tabularx}
 \caption{\label{tab:cu20_param}Parameters of the first peak of the Cu-O pRDF for Cu$_2$O obtained by sparse modeling (SpM) and by peak-fitting scheme using GNXAS for the same experimental data. Values taken from references are also shown for comparison.}
  \end{table}

\section{Conclusions}

We applied sparse modeling to the analysis of EXAFS data through the use of LASSO. Starting from the $n$-body expansion, the signal is described by the two-body term and the appropriate value for the hyper-parameter $\lambda$ is chosen by cross-validation analysis and the 1SE rule. We tested the method on different systems. Without using any prior information about the structure, other than knowing which elements are present in the material, we were able to obtain peak positions of the RDF for metals and cuprous oxide, which is a step forward respect to previous formulation \cite{AkaIwa2018}. A quantitative analysis can then be performed on the nearest-neighbor peaks to retrieve structural parameters, like average distances and Debye-Waller factors. 

Respect to the conventional approach, this method is mostly automatic and easy to use, since very few inputs are required. Another advantage is also the time required to carry out the fit: on a standard laptop results are available within seconds, enabling the possibility to analyze large amount of data in a timely manner, or check the results at the same time of the experiment. Further improvements to the method are currently under investigation.

\section*{Funding}
This work is supported by JST CREST (JPMJCR1861, JPMJCR1761) and JST PRESTO (JPMJPR17N2).

\bibliography{biblio.bib}

\end{document}